\documentstyle[11pt,newpasp,twoside,epsf]{article}
\def\edcomment#1{\iffalse\marginpar{\raggedright\sl#1\/}\else\relax\fi}
\reversemarginpar

\begin{document}
\title{AGN-Galaxy Clustering with the Panoramic Deep Fields}
\author{Michael J.I. Brown}
\affil{National Optical Astronomy Observatory, 950 N. Cherry Ave., 
P.O. Box 26732, Tucson, Arizona 85726, U.S.A.}

\begin{abstract}
The AGN-galaxy cross-correlation function of radio-quiet AGN and 
radio galaxies has been measured with the Panoramic Deep Fields.
Colour selection criteria and photometric redshifts
have been used to significantly increase the signal-to-noise of the angular
cross-correlation function. Radio-quiet AGN environments are comparable
to the environments of early-type galaxies at low redshift.
The radio galaxy-galaxy spatial cross-correlation function is 
very strong though large variations are observed from field-to-field.
These variations appear to be caused by large-scale-structure 
on scales comparable to the $5^\circ \times 5^\circ$ 
field-of-view. The distribution and spatial cross-correlation 
function of radio galaxies and clusters in the Panoramic Deep 
Fields is consistent with these objects tracing the same 
structures at $z<0.7$. No evidence is found for evolution of the 
AGN-galaxy spatial cross-correlation function across the redshift range 
observed.

\end{abstract}

\section{The Panoramic Deep Fields}

The environments of radio-quiet AGN and radio galaxies have been 
measured using the Panoramic Deep Fields, a $UB_JRI$ imaging 
survey of two $5^\circ \times 5^\circ$ fields with a depth of 
$B_J\sim 23.5$ and $R\sim 22.5$. The wide field-of-view and 
depth was obtained by coadding SuperCOSMOS scans of UK Schmidt 
photographic plates. The resulting galaxy catalogues contain
more than $2\times 10^5$ galaxies per field and thus provide
the large sample size required for accurate estimates of AGN-galaxy 
clustering. As the galaxy catalogues contain more than 
1000 objects with spectroscopic redshifts from the NED database, 
the polynomial fitting
technique of Connolly {\it et al.} (1995) was used to estimate
photometric redshifts for all galaxies with $B_J$ and 
$R$-band detections. The B1950 coordinates of the field centres 
are $00~53~-28~03$ (SGP field) and $10~40~+00~00$ (F855 field).
Access to the galaxy catalogue and cutout images are available via 
{\tt http://astro.ph.unimelb.edu.au/data/}.

\section{Radio-quiet AGN}

The environments of $z<0.7$ radio-quiet AGN were measured using the 
Panoramic Deep Fields galaxy catalogues and 109 $UB_JR$ selected
AGN with spectroscopic identifications from La Franca {\it et al.} (1999) 
and Croom {\it et al.} (2001). Galaxy clustering is a function 
of restframe colour so the galaxy sample was split into red 
(early-type) and blue (late-type) subsamples which were 
selected using the $B_J-R$ colour of a non-evolving Sbc 
template as a function of $z$ and the galaxy photometric redshifts. 

AGN-galaxy clustering was measured using the angular cross-correlation 
function. The spatial cross-correlation function was then determined with 
Limber's equation (Limber 1954), the angular cross-correlation function, 
AGN spectroscopic redshifts and the smoothed distribution of 
galaxy photometric redshifts. A thorough description of 
the techniques and their application to the SGP field is 
provided by Brown, Boyle \& Webster (2001). The clustering
of blue galaxies around radio-quiet AGN is weak and was detected with 
low signal-to-noise resulting in poor constraints on the 
radio-quiet AGN-blue galaxy spatial cross-correlation function. 

The signal-to-noise of the angular cross-correlation function is decreased
by AGN-galaxy pairs which are not physically associated with each other.
For objects with good redshift estimates, it is possible to increase 
the signal-to-noise of the angular cross-correlation function by excluding 
object pairs which can not be physically associated. While the 
photometric redshifts of $z>0.3$ blue galaxies have errors comparable 
to the redshift estimate, the photometric redshifts of red galaxies 
have $1\sigma$ errors of $<20\%$. The radio-quiet AGN-red galaxy angular 
cross-correlation function determined with AGN-galaxy pairs with 
redshifts within $2\sigma$ of each other is shown in Figure~1. 
A power-law provides a good approximation of the observed clustering 
in both of the Panoramic Deep Fields. The spatial cross-correlation was approximated by a power-law 
of the form
\begin{equation}
\xi(r)=(r/r_0)^{-\gamma}
\end{equation}
where $r$ is the spatial separation in comoving coordinates and 
$r_0$ and $\gamma$ are constants. The values of $r_0$ (with $\gamma=1.9$) 
are $8\pm 3 h^{-1} {\rm Mpc}$ and $10\pm 2 h^{-1} {\rm Mpc}$ in the 
F855 and SGP fields respectively. The values of $r_0$ are comparable
to those determined for early-type galaxies (Brown, Boyle \& Webster 2001).
There is no evidence of a systematic increase or decrease of $r_0$ with 
redshift.

\begin{figure}
\vspace*{90pt}
\plotfiddle{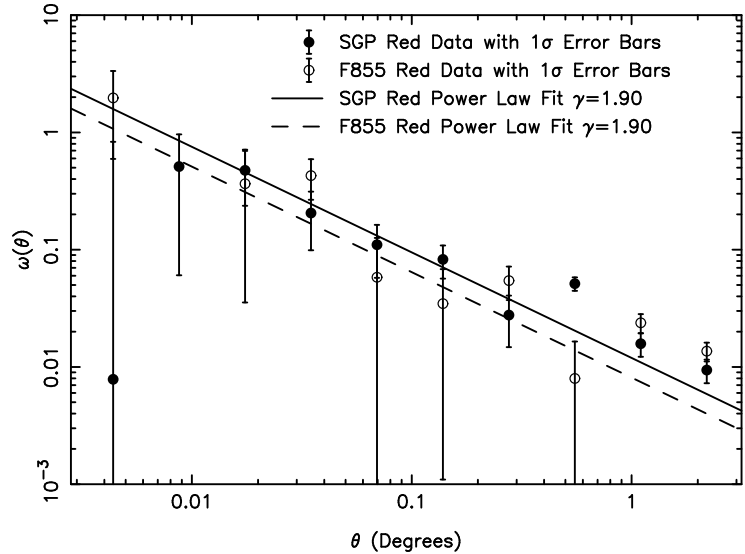}{90pt}{0}{50}{50}{-210}{-50}
\caption{The $UB_JR$ selected AGN-red galaxy angular cross-correlation 
function. The signal-to-noise has been increased by only including
AGN-galaxy pairs where the galaxy photometric redshifts are within
$2\sigma$ of the AGN spectroscopic redshifts.}
\end{figure}

\section{Radio Galaxies}

The environments of radio galaxies were measured using 
230 radio galaxies from Brown, Webster \& Boyle (2001). 
The radio galaxy-galaxy angular cross-correlation function 
is significantly stronger than the radio-quiet AGN-galaxy cross-correlation
function. However, estimates of the value of $r_0$ vary by 
a factor of $\sim 2$ between the SGP and F855 fields. 
A possible cause of the variations from field-to-field is 
large-scale-structure. The cluster catalogue of 
Brown, Webster \& Boyle (2002) shows structures with sizes comparable
to the field-of-view. Figure~2, a comparison of the distribution of 
radio galaxies and clusters in the F855 field, shows radio galaxies
and clusters tracing the same large-scale-structure across the field.

The radio galaxy-cluster spatial cross-correlation function is
extremely strong with $\gamma\sim 2.8$ and $r_0$ values of 
$9\pm 2 h^{-1} {\rm Mpc}$ and $10\pm 2 h^{-1} {\rm Mpc}$ in the 
F855 and SGP fields respectively. The strength of the clustering
is consistent with radio galaxies tracing the same structures
as clusters. As with radio-quiet AGN, there is no evidence of 
significant evolution of $r_0$ with redshift. 

\begin{figure}
\plotone{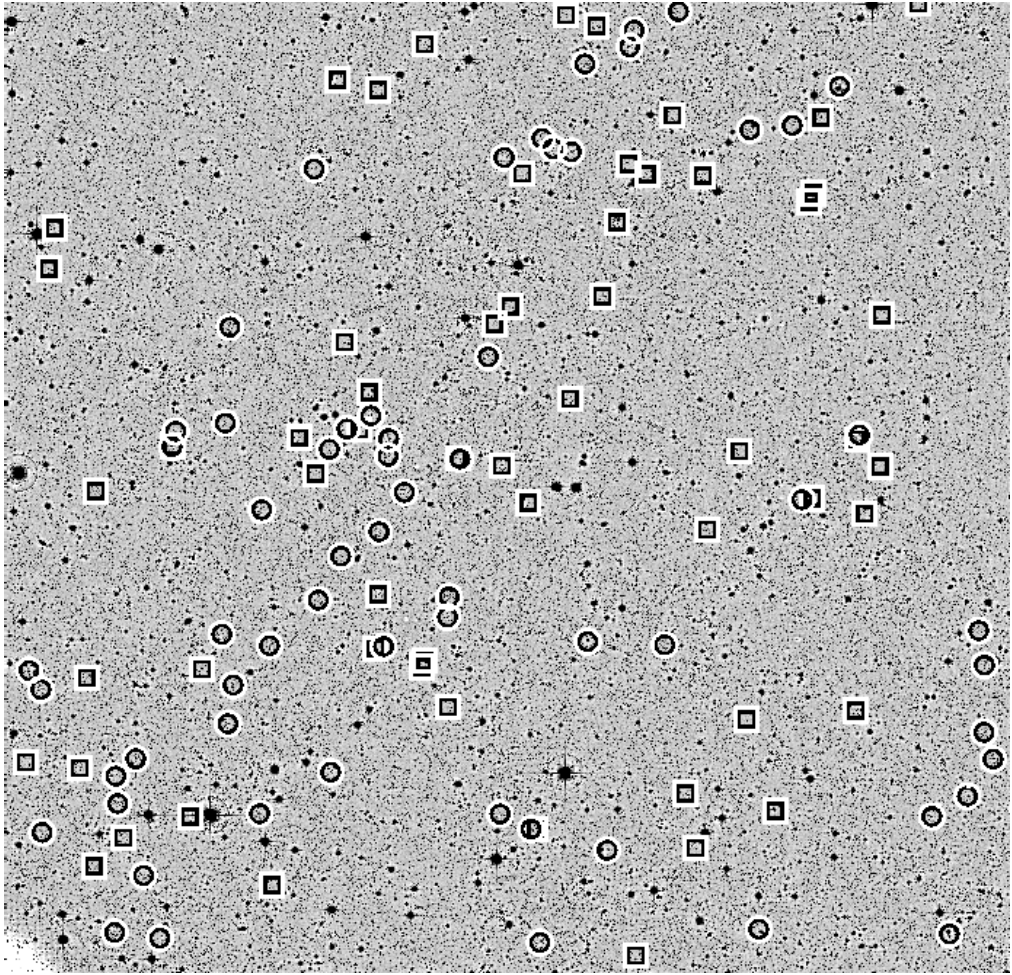}
\caption{The $R$-band image of the F855 field with the
distribution of radio galaxies (squares) and galaxy cluster candidates
(circles) with photometric redshifts in the range $0.3<z<0.5$.
Except for regions near bright stars, the selection functions for 
radio galaxies and galaxy clusters are uniform across the field-of-view.
Several voids are apparent in the distribution of clusters and radio 
galaxies. At $z\sim 0.4$, $5^\circ$
is equivalent to $\sim 100 h^{-1} {\rm Mpc}$ (comoving) transverse to the line-of-sight.}
\end{figure}

\section{Summary}

The Panoramic Deep Fields have been used to measure the 
AGN-galaxy cross-correlation function for radio-quiet AGN and
radio galaxies. By applying colour selection and photometric 
redshift criteria, it has been possible to significantly increase 
the signal-to-noise of the angular cross-correlation function. 
The clustering of red (early-type) galaxies around $UB_JR$ selected 
AGN is comparable to the clustering of early-type galaxies at low 
redshift. The clustering of galaxies around radio galaxies is strong
though large variations in the clustering strength are observed between 
the two Panoramic Deep Fields. The distribution and spatial cross-correlation function of 
radio galaxies and galaxy clusters indicate that these objects trace
the same large-scale-structures.

\end{document}